\documentclass[preprint,eqsecnum,preprintnumbers,nofootinbib,byrevtex,prd,aps,showpacs,showkeys,groupedaddress,floatfix]{revtex4}
\usepackage{bm}
\usepackage{graphics}
\usepackage{graphicx}
\usepackage{epsfig}
\usepackage{amssymb}
\usepackage{amsmath}
\begin{document}

\title{Analytical solutions of the $D$-dimensional Schr\"{o}dinger
equation   with the Woods-Saxon potential for arbitrary $l$ state}
\author{V.~H.~Badalov$^{1}$} \email{E-mail:badalovvatan@yahoo.com}%
\author{H.~I.~Ahmadov$^{2}$} \email{E-mail:hikmatahmadov@yahoo.com}
\affiliation{$^{1}$Institute for Physical Problems, Baku State
University, Z. Khalilov st. 23, AZ-1148, Baku, Azerbaijan  \\
$^{2}$Department of Mathematical Physics\\Faculty of Applied
Mathematics and Cybernetics, Baku State University, Z. Khalilov
st. 23, AZ-1148, Baku, Azerbaijan}


\begin{abstract}
{In this work, the analytical solution of the hyper-radial
Schr\"{o}dinger equation for the spherical Woods-Saxon potential
in D dimensions is presented. In our calculations, we have applied
the Nikiforov-Uvarov method by using the Pekeris approximation to
the centrifugal potential for arbitrary $l$ states. The bound
state energy eigenvalues and corresponding eigenfunctions are
obtained for various values of $n$ and $l$ quantum numbers.}
\end{abstract}


\maketitle

\section{\bf Introduction}

An analytical solution of the radial Schr\"{o}dinger equation is
of high importance in nonrelativistic quantum mechanics, because
the wave function contains all necessary information for full
description of a quantum system. There are only few potentials for
which the radial Schr\"{o}dinger equation can be solved explicitly
for all $n$ and $l$ quantum numbers. So far, many methods were
developed, such as supersymmetry (SUSY) [1,2] and the Pekeris
approximation [3-8], to solve the radial Schr\"{o}dinger equation
exactly or quasi-exactly for $l\neq0$ within these potentials.
Levai and Williams suggested a simple method for constructing
potentials for which the Schr\"{o}dinger equation can be solved
exactly in terms of special functions [9] and showed relationship
between the introduced formalism  and supersymmetric quantum
mechanics [1]. Amore \emph{et al.} presented a new method [10] for
the solution of the Schr\"{o}dinger equation applicable to
problems of nonperturbative nature. In addition, they applied the
method to the quantum anharmonic oscillator and found energy
eigenvalues and wave functions, even for strong couplings.

The three-dimensional radial Schr\"{o}dinger equation for the
sperical Woods-Saxon potential [11] can not be solved analytically
for $l\neq0$ states, because of the centrifugal term $\sim
r^{-2}$. However, Fl\"{u}gge gave an exact expression for the wave
function, but a graphical method was suggested for the energy
eigenvalues at $l=0$ [4]. It is well known that the Woods-Saxon
potential is one of the important short-range potentials in
physics. Furthermore, this potential were applied to numerous
problems, in nuclear and particle physics, atomic physics,
condensed matter, and chemical physics.

Recently, an alternative method known as the Nikiforov-Uvarov (NU)
method [12] was proposed for solving the Schr\"{o}dinger equation.
Therefore, it would be interesting and important to solve the
nonrelativistic radial Schr\"{o}dinger equation for Woods-Saxon
potential for $l\neq0$, since it has been extensively used to
describe the bound and continuum states of the interacting systems.
Thus, one needs to obtain the energy eigenvalues and corresponding
eigenfunctions of the one particle problem within this potential.
The NU method was used by C. Berkdemir et al. [13] to solve the
radial Schr\"{o}dinger equation for the generalized Woods-Saxon
potential for $l=0$. However, it this work, the authors made errors
in application of the NU method, which led to no-correct results
[14]. In Refs.[7,8] the radial Schr\"{o}dinger equation with the
spherical Woods-Saxon potential have been solved using the
Nikiforov-Uvarov method and Pekeris approximation for arbitrary $l$
states. The same method has been employed to solve the Klein-Gordon
equation with the spherical Woods-Saxon potential for $l\neq0$
states Ref.[15].

In this work, we solve  the D dimensional hyper-radial
Schr\"{o}dinger equations with spherical Woods-Saxon potential for
arbitrary $l$ states. Here the NU method [12] and the Pekeris
approximattion is applied to find energy eigenvalues and
corresponding eigenfunctions of the considering problem.

\section{\bf Nikiforov-Uvarov method}

The Nikiforov-Uvarov (NU) method is based on the solutions of
general second-order linear equations with special orthogonal
functions. It has been extensively used to solve the nonrelativistic
Schr\"{o}dinger equation and other Schr\"{o}dinger-like equations.
The one-dimensional Schr\"{o}dinger equation or similar second-order
differential equations can be written using NU method in the
following form:
\begin{equation}
\psi''(z)+\frac{\widetilde{\tau}(z)}{\sigma(z)}{\psi}'(z)+
\frac{\widetilde{\sigma}(z)}{\sigma^{2}(z)}\psi(z)=0,
\end{equation}
where $\sigma(z)$ and $\widetilde{\sigma}(z)$ are polynomials, at
most second-degree, and $\widetilde{\tau}(z)$ is a first-degree
polynomial.

Using Eq.(2.1) the transformation
\begin{equation}
\psi(z)=\Phi(z)\\{y}(z)
\end{equation}
one reduces it to the hypergeometric-type equation
\begin{equation}
\sigma(z){y}''+\tau(z){y}'+\lambda{y}=0.
\end{equation}
The function $\Phi(z)$ is defined as the logarithmic derivative [12]
\begin{equation}
\frac{\Phi'(z)}{\Phi(z)}=\frac{\pi(z)}{\sigma(z)},
\end{equation}
where $\pi(z)$ is at most the first-degree polynomial.

The another part of $\psi(z)$, namely ${y}(z)$, is the
hypergeometric-type function, that for fixed $n$ is given by the
Rodriguez relation:
\begin{equation}
{y_{n}}(z)=\frac{{B_{n}}}{\rho(z)}\frac{{d^{n}}}{{dz^{n}}}[\sigma^{n}(z)\rho(z)],
\end{equation}
where ${B_{n}}$ is the normalization constant and the weight
function $\rho(z)$ must satisfy the condition [12]
\begin{equation}
\frac{d}{dz}\left(\sigma(z)\rho(z)\right)=\tau(z)\rho(z),
\end{equation}
with $\tau(z)=\widetilde{\tau}(z)+2\pi(z).$

For accomplishment of the conditions imposed on function $\rho(z)$,
the classical orthogonal polynomials, it is necessary, that
polynomial $\tau(z)$ becomes equal to zero in some point of an
interval $(a,b)$ and derivative of this polynomial for this interval
at $\sigma(z)>0$ will be negative, i.e. $\tau'(z)<0$.

The function $\pi(z)$ and the parameter $\lambda$ required for this
method are defined as follows:
\begin{equation}
\pi(z)=\frac{\sigma'-\widetilde{\tau}}{2}\pm\sqrt{\left(\frac{\sigma'-
\widetilde{\tau}}{2}\right)^{2}-\widetilde{\sigma}+{k}\sigma},
\end{equation}
\begin{equation}
\lambda=k+\pi'(z).
\end{equation}
On the other hand, in order to find the value of $k$, the expression
under the square root must be the square of a polynomial. This is
possible only if its discriminant is zero. Thus, the new eigenvalue
equation for the Schr\"{o}dinger equation becomes [12]
\begin{equation}
\lambda=\lambda_{n}=-n\tau'-\frac{n(n-1)}{2}\sigma'', (n=0,1,2,...).
\end{equation}
After the comparison of Eq.(2.8) with Eq.(2.9), we obtain the energy
eigenvalues.

\section{\bf Solutions of the Schr\"{o}dinger equation with the
Woods-Saxon potential}

Woods and Saxon introduced a potential to study elastic scattering
of 20 MeV protons by a heavy nuclei [11]. The spherical Woods-Saxon
potential that was used as a major part of nuclear shell model, has
received a lot of attention in nuclear mean field model. The
spherical standard Woods-Saxon potential [11] is defined by

\begin{equation}
V(r)=-\frac{V_{0}}{1+\exp\left(\frac{r-R_{0}}{a}\right)}, a<<R_{0}.
\end{equation}
This potential was used for description of interaction of a neutron
with a heavy nucleus. The parameter $R_{0}$ is interpreted as radius
of a nucleus, the parameter $a$ characterizes thickness of the
superficial layer inside, which the potential falls from value $V=0$
outside of a nucleus up to value $V=-V_{0}$ inside a nucleus. At
$a=0$, one gets the simple potential well with jump of potential on
the surface of a nucleus.

Using $D$-dimensional ($D\geq2$) polar coordinates with polar
variable $r$ (hyperradius) anf angular variables $\theta_{1},
\theta_{2}, \ldots , \theta_{D-2}, \phi$ (hyperangles), the
Laplasian operator in polar coordinates $(r, \theta_{1}, \theta_{2},
\ldots , \theta_{D-2}, \phi )$ of $R^{D}$ is
$$\nabla_{D}^{2}=r^{1-D}\frac{\partial}{\partial
r}\left(r^{D-1}\frac{\partial}{\partial
r}\right)+\frac{\Lambda_{D}^{2}}{r^{2}},$$
where $\Lambda_{D}^{2}$
is a partial differential operator on the unit sphere $S^{D-1}$
(Laplace-Beltrami operator, or grand orbital operator, or
hyperangular momentum operator) defined analogously to a
three-dimensional angular momentum, Avery [16].

The $D$ dimensional  Schr\"{o}dinger equation with spherically
symmetric potential $V(r)$ has the form [16]
\begin{equation}
\left(-\frac{\hbar^{2}}{2\mu}\nabla_{D}^{2}+V(r)-E_{nl}\right)\psi_{nlm}(r,\Omega_{D})=0,
\end{equation}
where $\mu$ is the reduced mass, $\hbar$ is the Planck's constant
and
\begin{equation}
\psi_{nlm}(r,\Omega_{D})=R_{nl}(r)Y_{lm}(\Omega_{D}).
\end{equation}

The Laplasian  operator divides into a hyper-radial part
$r^{1-D}\frac{\partial}{\partial
r}\left(r^{D-1}\frac{\partial}{\partial r}\right)$ and an angular
part
$\frac{\Lambda_{D}^{2}}{r^{2}}=-\frac{\widehat{L}_{D}^{2}}{\hbar^{2}r^{2}}$,
i.e.
\begin{equation}
\nabla_{D}^{2}=r^{1-D}\frac{\partial}{\partial
r}\left(r^{D-1}\frac{\partial}{\partial
r}\right)-\frac{\widehat{L}_{D}^{2}}{\hbar^{2}r^{2}},
\end{equation}
where $\widehat{L}_{D}$ is the grand orbital angular momentum
operator. The eigenfunctions of $\widehat{L}_{D}^{2}$ are the
hyper-spherical harmonics $Y_{lm}(\Omega_{D})$
\begin{equation}
L_{D}^{2}Y_{lm}(\Omega_{D})=\hbar^{2}l(l+D-2)Y_{lm}(\Omega_{D}),
\end{equation}
where $l$ is the angular momentum quantum number.

After substituting the Eqs.(3.3), (3.4), (3.5) into (3.2) and using
the fact that $\psi_{nlm}(r,\Omega_{D})$ is the eigenfunction of
$\widehat{L}_{D}^{2}$ with eigenvalue $\hbar^{2}l(l+D-2)$, we obtain
an equation known as the hyper-radial Schr\"{o}dinger equation with
Woods-Saxon potential
\begin{equation}
\frac{d^2R_{nl}(r)}{dr^2}+\frac{D-1}{r}\frac{dR_{nl}(r)}{dr}+\frac{2\mu}{\hbar^{2}}\left[
E+\frac{V_0 }{1+\exp \left(
\frac{r-R_0}{a}\right)}-\frac{\hbar^{2}l(l+D-2)}{2\mu
r^2}\right]R_{nl}(r)=0,  (0\leq r<\infty).
\end{equation}

Introducing a new function
$$u_{nl}(r)=r^{\frac{D-1}{2}}R_{nl}(r),$$
Eq.(3.6) reduces to

\begin{equation}
\frac{d^2u_{nl}(r)}{dr^2}+\frac{2\mu}{\hbar^2}\left[E+\frac{V_0}{1+\exp\left(\frac{r-R_0}{a}\right)}-
\frac{\hbar^{2}\left(l+\frac{D-1}{2}\right)\left(l+\frac{D-3}{2}\right)}{2\mu
r^2}\right]u_{nl}(r)=0.
\end{equation}
introducing a new parametr
$$\widetilde{l}=l+\frac{D-3}{2},$$
Eq.(3.7) takes the form
\begin{equation}
\frac{d^2u_{nl}(r)}{dr^2}+\frac{2\mu}{\hbar^2}\left[E+\frac{V_0}{1+\exp\left(\frac{r-R_0}{a}\right)}-
\frac{\hbar^{2}\widetilde{l}\left(\widetilde{l}+1\right)}{2\mu
r^2}\right]u_{nl}(r)=0.
\end{equation}
Equation (3.8) has the same form as the equation for a particle in
one dimension, except for two important differences. First, there is
a repulsive effective potential proportional to the eigenvalue of
$\hbar^{2}\widetilde{l}(\widetilde{l}+1)$. Second, the radial
function must satisfy the boundary conditions $u(0)=0$ and
$u(\infty)=0.$

It is sometimes convenient to define in Eq.(3.8) the effective
potential in the form:
\begin{equation}
V_{eff}(r)=V(r)+\frac{\hbar^{2}\widetilde{l}(\widetilde{l}+1)}{2\mu
r^{2}}.
\end{equation}

Then, the radial Schr\"{o}dinger equation  given by Eq.(3.8) takes
the form
\begin{equation}
\frac{d^2u_{nl}(r)}{dr^2}+\frac{2\mu }{\hbar^{2}} \left[
E-V_{eff}(r)\right]u_{nl}(r)=0.
\end{equation}

If in Eq.(3.1) introduce the notations
$$x=\frac{r-R_{0}}{R_{0}},\,\,\, \alpha=\frac{R_{o}}{a},$$
then the Woods-Saxon potential is given by the expression
$$V_{WS}=-\frac{V_{0}}{1+\exp(\alpha x)}.$$
The effective potential together with the WS potential for $l\neq0$
can be written as
\begin{equation}
V_{eff}(r)=V_{l}(r)+V_{WS}(r)=\frac{\hbar^{2}\widetilde{l}(\widetilde{l}+1)}{2\mu
r^{2}}-\frac{V_{0}}{1+\exp(\alpha x)}.
\end{equation}

It is known that the Schr\"{o}dinger equation cannot be solved
exactly for this potential at the value $\widetilde{l}\neq0$ using
the standard methods as SUSY and NU. From Eq.(3.11) it is seen that
the effective potential is a combination of the exponential and
inverse square potentials, which cannot be solved analytically.
Therefore, in order to solve this problem we can take the most
widely used and convenient for our purposes Pekeris approximation
[3-8, ]. This approximation is based on the expansion of the
centrifugal barrier in a series of exponentials depending on the
internuclear distance, taking into account terms up to second order,
so that the effective $l$-dependent potential preserves the original
form. It should be pointed out, however, that this approximation is
valid only for low vibrational energy states. By changing the
coordinates $x=\frac{r-R_{0}}{R_{0}}$ or $r=R_{0}(1+x)$, the
centrifugal potential is expanded in the Taylor series around the
point $x=0$ ($r=R_{0}$)
\begin{equation}
V_{l}(r)=\frac{\hbar^{2}\widetilde{l}(\widetilde{l}+1)}{2\mu
r^{2}}=\frac{\hbar^{2}\widetilde{l}(\widetilde{l}+1)} {2\mu
R_{0}^{2}}\frac{1}{(1+x)^{2}}=\delta\left(1-2x+3x^{2}-4x^{3}+\ldots\right),
\end{equation}
where
$\widetilde{\delta}=\frac{\hbar^{2}\widetilde{l}(\widetilde{l}+1)}{2\mu
R_{0}^{2}}$.

According to the Pekeris approximation, we shall replace potential
$V_{\widetilde{l}}(r)$ with expression

\begin{equation}
V^{*}_{\widetilde{l}}(r)=\widetilde{\delta}
\left(C_{0}+\frac{C_{1}}{1+\exp\alpha
x}+\frac{C_{2}}{\left(1+\exp\alpha x\right)^{2}}\right).
\end{equation}

In order to define the parameters $C_{0}$, $C_{1}$ and $C_{2}$, we
also expand this potential in the Taylor series around the point
$x=0$ ($r=R_{0}$):
\begin{equation}
V^{*}_{\widetilde{l}}(x)=\widetilde{\delta}\left[\left(C_{0}+\frac{C_{1}}{2}+\frac{C_{2}}{4}\right)-
\frac{\alpha}{4}\left(C_{1}+C_{2}\right)x+\frac{\alpha^{2}}{16}C_{2}x^{2}+\frac{\alpha^{3}
}{48}\left(C_{1}+C_{2}\right)x^{3}-\frac{\alpha^{4}}{96}C_{2}x^{4}+\cdots\right].
\end{equation}
Comparing equal powers of $x$ Eqs.(3.12) and (3.14), we obtain the
constants $C_{0}$, $C_{1}$ and $C_{2}$:
$$C_{0}=1-\frac{4}{\alpha}+\frac{12}{\alpha^{2}},\,\,
C_{1}=\frac{8}{\alpha}-\frac{48}{\alpha^{2}},\,\,
C_{2}=\frac{48}{\alpha^{2}}.$$

Now, the effective potential after Pekeris approximation becomes
equal to
\begin{equation}
V^{*}_{eff}(x)=V^{*}_{l}(x)+V_{WS}(x)=\widetilde{\delta}
C_{0}-\frac{V_{0}-\widetilde{\delta} C_{1}}{1+\exp(\alpha x)}+
\frac{\widetilde{\delta} C_{2}}{\left(1+\exp(\alpha
x)\right)^{2}}.
\end{equation}

Instead of solving the radial Schr\"{o}dinger equation for the
effective Woods-Saxon potential $V_{eff}(r)$  given by Eq.(3.11), we
now solve the radial Schr\"{o}dinger equation for the new effective
potential $V^{*}_{eff}(r)$ given by Eq.(3.15) obtained using the
Pekeris approximation. Having inserted this new effective potential
into Eq.(3.10), we obtain
\begin{equation}
\frac{d^2u_{nl}(r)}{dr^2}+\frac{2\mu }{\hbar^{2}} \left[
E-\widetilde{\delta} C_{0}+\frac{V_{0}-\widetilde{\delta}
C_{1}}{1+\exp\left(\frac{r-R_{0}}{a}\right)}-\frac{\widetilde{\delta}
C_{2}}{\left(1+\exp\left(\frac{r-R_{0}}{a}\right)\right)^{2}}\right]u_{nl}(r)=0.
\end{equation}
We use the following dimensionless notations:
\begin{equation}
\epsilon^2=-\frac{2\mu \left(E-\widetilde{\delta}
C_{0}\right)a^2}{\hbar ^2};\,\,\beta^2=\frac{2\mu
\left(V_{0}-\widetilde{\delta} C_{1}\right)a^2}{\hbar ^2};\,\,
\gamma^{2}=\frac{2\mu\widetilde{\delta} C_{2}a^2}{\hbar^2},
\end{equation}
with  real $\epsilon>0$ for bound states; $\beta$ and $\gamma$ are
real and positive.

If we rewrite Eq.(3.16) by using a new variable of the form
$$z=\left(1+\exp\left(\frac{r-R_{0}}{a}\right)\right)^{-1},$$ we obtain
\begin{equation}
u^{\prime \prime }(z)+\frac{1-2z}{z(1-z)}u^{\prime }(z)+
\frac{-\epsilon^{2}+\beta^{2}z-\gamma^{2}z^{2}}{(z(1-z))^{2}}u(z)=0,
(0\leq z\leq 1),
\end{equation}
with $\widetilde{\tau}(z)=1-2z;  \sigma (z)=z(1-z);
\widetilde{\sigma}(z)=-\epsilon^2+\beta^2z-\gamma^{2}z^{2}$.

In the NU-method, the new function $\pi(z)$ is
\begin{equation}
\pi(z)=\pm\sqrt{\epsilon^2+\left(k-\beta^2\right)z-\left(k-\gamma^{2}\right)z^2}.
\end{equation}

The constant parameter $k$ can be found employing the condition that
the expression under the square root has a double zero, i.e., its
discriminant is equal to zero. Hence, there are two possible
functions for each $k$:
\begin{equation}
\label{2}\pi(z)=\pm\left\{
\begin{array}{c}
\left(\epsilon-\sqrt{\epsilon^{2}-\beta^{2}+\gamma^{2}}\right)z-\epsilon,\,\,\,
for \,\,\,
k=\beta^{2}-2\epsilon^{2}+2\epsilon\sqrt{\epsilon^{2}-\beta^{2}+\gamma^{2}},\\
\left(\epsilon+\sqrt{\epsilon^{2}-\beta^{2}+\gamma^{2}}\right)z-\epsilon,\,\,\,
for \,\,\,\,
k=\beta^{2}-2\epsilon^{2}-2\epsilon\sqrt{\epsilon^{2}-\beta^{2}+\gamma^{2}}.\,
\end{array}
\right.
\end{equation}
According to the NU-method, from the four possible forms of the
polynomial $\pi(z)$ we select the one for which the function
$\tau(z)$ has the negative derivative and root lies in the interval
(0,1). Therefore, the appropriate functions $\pi(z)$ and $\tau(z)$
have the following forms:
\begin{equation}
\pi(z)=\epsilon-\left(\epsilon+\sqrt{\epsilon^{2}-\beta^{2}+\gamma^{2}}\right)z,
\end{equation}

\begin{equation}
\tau(z)=1+2\epsilon-2\left(1+\epsilon+\sqrt{\epsilon^{2}-\beta^{2}+\gamma^{2}}\right)z,
\end{equation}
and
\begin{equation}
k=\beta^{2}-2\epsilon^{2}-2\epsilon\sqrt{\epsilon^{2}-\beta^{2}+\gamma^{2}}.
\end{equation}
Then, the constant $\lambda=k+\pi'(z)$ is written as
\begin{equation}
\lambda=\beta^{2}-2\epsilon^{2}-2\epsilon\sqrt{\epsilon^{2}-\beta^{2}+\gamma^{2}}-
\epsilon+\sqrt{\epsilon^{2}-\beta^{2}+\gamma^{2}}.
\end{equation}
An alternative definition of $\lambda_n$ (Eq.(2.9)) is
\begin{equation}
\lambda=\lambda_n=2\left(\epsilon+\sqrt{\epsilon^{2}-
\beta^{2}+\gamma^{2}}\right)n+n(n+1).
\end{equation}
Having compared Eqs.(3.19) and (3.20)
\begin{equation}
\beta^{2}-2\epsilon^{2}-2\epsilon\sqrt{\epsilon^{2}-\beta^{2}+\gamma^{2}}-
\epsilon-\sqrt{\epsilon^{2}-\beta^{2}+\gamma^{2}}=2\left(\epsilon+\sqrt{\epsilon^{2}-
\beta^{2}+\gamma^{2}}\right)n+n(n+1),
\end{equation}
we obtain
\begin{equation}
\epsilon+\sqrt{\epsilon^{2}-\beta^{2}+\gamma^{2}}+
n+\frac{1}{2}-\frac{1}{2}\sqrt{1+4\gamma^{2}}=0,
\end{equation}
or
\begin{equation}
\epsilon+\sqrt{\epsilon^{2}-\beta^{2}+\gamma^{2}}= n'.
\end{equation}
After some elementary calculus, one sees that
\begin{equation}
\epsilon-\sqrt{\epsilon^{2}-\beta^{2}+\gamma^{2}}=
\frac{\beta^{2}-\gamma^{2}}{n'}.
\end{equation}
Here
\begin{equation}
n'=-n+\frac{\sqrt{1+4\gamma^{2}}-1}{2},
\end{equation}
$n$ being the radial quantum number $(n=0,1,2,\ldots)$. From
Eq.(3.28) and (3.29), we find
\begin{equation}
\epsilon=\frac{1}{2}\left(n'+\frac{\beta^{2}-\gamma^{2}}{n'}\right)
\end{equation}
and
\begin{equation}
\sqrt{\epsilon^{2}-\beta^{2}+\gamma^{2}}=\frac{1}{2}\left(n'-\frac{\beta^{2}-\gamma^{2}}{n'}\right)
\end{equation}
Because for the bound states $\epsilon>0$ and Eqs.(3.28, 3.29, 3.31,
3.32) we get
\begin{equation}
n'>0
\end{equation}
and
\begin{equation}
0<\beta^{2}-\gamma^{2}<n'^{2}.
\end{equation}
If $n'>0$, there exist bound states, otherwise, there are no bound
states at all. By using Eq.(3.30) this relation can be recast into
the form
\begin{equation}
0\leq n<\frac{\sqrt{1+4\gamma^{2}}-1}{2},
\end{equation}
i.e. it gives the finite coupling value.

If $\beta^{2}-\gamma^{2}>0$, there exists bound states; otherwise
there are no bound states. Inequality, which obtained after
substituting $\beta, \gamma, C_{1}, C_{2}$ into Eq.(3.34), gives the
definite coupling value for the potential depth $V_{0}$:
\begin{equation}
V_{0}>\frac{\hbar^2 \widetilde{l}(\widetilde{l}+1)a}{2 \mu
R^{3}_{0}}
\end{equation}

After substituting $\gamma, \widetilde{\delta}, C_{2}$ into
Eq.(3.36), we find
\begin{equation}
0\leq n<\frac{\sqrt{1+\frac{192a^4l(l+1)}{R_{0}^4}}-1}{2}
\end{equation}

 The exact energy eigenvalues of the
Schr\"{o}dinger equation with the Woods-Saxon potential are derived
as
\begin{equation}
E_{nl}^{(D)}=\widetilde{\delta} C_{0}-\left(V_{0}-\widetilde{\delta}
C_{1}\right)\left(\frac{n'^2+\beta^2-\gamma^{2}}{2\beta
n'}\right)^{2}.
\end{equation}
Substituting the values of $\widetilde{\delta}, C_{0}, C_{1}, C_{2},
n', \beta$ and $\gamma$ into (3.38), one can find $E_{nl}$
$$E_{nl}^{(D)}=\frac{\hbar^2 \widetilde{l}(\widetilde{l}+1)}{2\mu
R_{0}^2}\left(1+\frac{12a^{2}}{R_{0}^2}\right)-$$
\begin{equation}
\frac{\hbar^2}{2\mu
a^2}\left\{\frac{\left[\sqrt{1+\frac{192\widetilde{l}(\widetilde{l}+1)a^{4}}{R_{0}^4}}-2n-1\right]^2}{16}+
\frac{4\left[\frac{\mu a^2V_{0}}{\hbar^2}
-\frac{4\widetilde{l}(\widetilde{l}+1)a^{3}}{R_{0}^3}\right]^2}{\left[\sqrt{1+\frac{192\widetilde{l}(\widetilde{l}+1)a^{4}}{R_{0}^4}}-2n-1\right]^2}+\frac{\mu
V_{0}a^2}{\hbar ^2}\right\}
\end{equation}

In three-dimensions,  the case when we set the parameter value
$D=3$ in Eq.(3.39), then we obtain the energy of the Woods-Saxon
potential
$$E_{nl}=\frac{\hbar^2 l(l+1)}{2\mu
R_{0}^2}\left(1+\frac{12a^{2}}{R_{0}^2}\right)-$$
\begin{equation}
\frac{\hbar^2}{2\mu
a^2}\left\{\frac{\left[\sqrt{1+\frac{192l(l+1)a^{4}}{R_{0}^4}}-2n-1\right]^2}{16}+
\frac{4\left[\frac{\mu a^2V_{0}}{\hbar^2}
-\frac{4l(l+1)a^{3}}{R_{0}^3}\right]^2}{\left[\sqrt{1+\frac{192l(l+1)a^{4}}{R_{0}^4}}-2n-1\right]^2}+\frac{\mu
V_{0}a^2}{\hbar ^2}\right\}
\end{equation}
which is identical to the one obtained in Ref.[8].

If both conditions (3.36) and (3.37)  are satisfied simultaneously,
the bound states exist. For very large $V_{0}$  the $l$-dependent
effective potential has the same form as the potential with $l=0$.
Whenever $D=3$, the bound states of the system do not exist in $l=0$
state, because (3.36) and (3.37) inequalities are not satisfied.
Hence, the radial Schr\"{o}dinger equation for the standard
Woods-Saxon potential with zero angular momentum has no bound state.
Furthermore, whenever $D>3$, the bound states of the system exist in
 $l=0$ state. Thus, the energy spectrum equation (3.39) is limited,
i.e. we have only the finite number of energy eigenvalues.

According to Eq.(3.39), the energy eigenvalues  depend on the depth
of the potential $V_{0}$, the width of the potential $R_{0}$,
surface thickness $a$, and the parameter $D$. If constraints imposed
on $n$ and $V_{0}$ satisfied, the bound states appear. From
Eq.(3.36), it is seen that the potential depth decreases when the
parameter $R_{0}$ increases and the parameter $a$ decreases, and
vice versa. Therefore, one can say that the bound states exist
within this potential.

In addition, we have seen that there are some restrictions on the
potential parameters for the bound state solutions within the
framework of quantum mechanics. That is, when the values of the
parameters $V_{0}$ and $n$ satisfy the conditions (3.36) and (3.37),
we obtain the bound states. We also point out that the exact results
obtained for the standard Woods-Saxon potential may have some
interesting applications for studying different quantum mechanical
and nuclear scattering problems. Consequently, the found wave
functions are physical ones.

Now, we are going to determine the radial eigenfunctions of this
potential. Having substituted $\pi(z)$ and $\sigma(z)$ into Eq.(2.4)
and then solving first-order differential equation, one can find the
finite function $\Phi(z)$ in the interval $[0,1]$

\begin{equation}
\Phi(z)=z^{\epsilon}\left(1-z\right)^{\sqrt{\epsilon^{2}-\beta^{2}+\gamma^{2}}}.
\end{equation}

It is easy to find the second part of the wave function from the
definition of weight function:
\begin{equation}
\rho(z)=z^{2\epsilon}\left(1-z\right)^{2\sqrt{\epsilon^{2}-\beta^{2}+\gamma^{2}}},
\end{equation}
and substituting into Rodrigues relation (2.4), we get
\begin{equation}
y_{n}(z)=B_{n}z^{-2\epsilon}\left(1-z\right)^{-2\sqrt{\epsilon^{2}-\beta^{2}+\gamma^{2}}}
\frac{d^{n}}{dz^{n}}\left[z^{n+2\epsilon}\left(1-z\right)^{n+2\sqrt{\epsilon^{2}-\beta^{2}+\gamma^{2}}}\right],
\end{equation}
where $B_{n}$ is the normalization constant and its value is
$\frac{1}{n!}$ [17]. Then, $y_{n}$ is given by the Jacobi
polynomials:
$$y_{n}(z)=P_{n}^{\left(2\epsilon,2\sqrt{\epsilon^{2}-\beta^{2}+\gamma^{2}}
\right)}(1-2z),$$
where $$P_{n}^{(\alpha,\beta)}(1-2z)=\frac{1}{n!}
z^{-\alpha}\left(1-z\right)^{-\beta}\frac{d^{n}}{dz^{n}}
\left[z^{n+\alpha}\left(1-z\right)^{n+\beta}\right].$$

The corresponding $u_{nl}(z)$ radial wave functions are found to be
\begin{equation}
u_{nl}(z)=C_{nl}z^{\epsilon}\left(1-z\right)^{\sqrt{\epsilon^{2}-\beta^{2}+\gamma^{2}}}P_{n}^{\left
(2\epsilon,2\sqrt{\epsilon^{2}-\beta^{2}+\gamma^{2}}\right)}(1-2z),
\end{equation}
where $C_{nl}$ is a new normalization constant determined using
$\int_o^\infty[u_{nl}(r)]^2dr=1$.

\section{\bf Conclusion}

In this paper, we have analytically calculated energy eigenvalues of
the bound states and corresponding eigenfunctions in the new exactly
solvable Woods-Saxon potential. The energy eigenvalue expression for
Woods-Saxon potentials is given by Eq.(3.39). As it should be
expected (see Eq.(3.39)), for any given set of parameters $V_{0},
R_{0}$, $a$ and $D$, the energy levels of standard Woods-Saxon
potential are positive. We can conclude that our results are
interesting not only for pure theoretical physicist but also for
experimental physicist, because the results are exact and more
general and useful to study nuclear scattering.

\end{document}